\documentclass[english,superscriptaddress,twocolumn,showpacs,pra]{revtex4-1}
\usepackage[T1]{fontenc}
\usepackage[latin9]{inputenc}
\usepackage{color}
\usepackage{amssymb}
\usepackage{graphicx}

\makeatletter
 \@ifundefined{textcolor}{}
 {%
   \definecolor{BLACK}{gray}{0}
   \definecolor{WHITE}{gray}{1}
   \definecolor{RED}{rgb}{1,0,0}
   \definecolor{GREEN}{rgb}{0,1,0}
   \definecolor{BLUE}{rgb}{0,0,1}
   \definecolor{CYAN}{cmyk}{1,0,0,0}
   \definecolor{MAGENTA}{cmyk}{0,1,0,0}
   \definecolor{YELLOW}{cmyk}{0,0,1,0}
 }

\makeatother

\usepackage{babel}
\begin{document}

\title{Decoherence via coupling to a finite quantum heat bath}

\author{O.Fialko}

\affiliation{Centre for Theoretical Chemistry and Physics, New Zealand Institute
for Advanced Study, Massey University (Albany Campus), Auckland, New
Zealand}
\begin{abstract}
Decoherence of a quantum state coupled to an exterior environment
is at the foundation of our understanding of the emergence of classical
behavior from the quantum world, but how does it emerge in a \textcolor{black}{finite
}closed quantum system? Here this is studied by modeling an isolated
quantum system of \textcolor{black}{a handful of} ultracold atoms
confined to a double well potential and coupled to a single atom of
a different type. \textcolor{black}{The ultracold atoms thermalize}
and serve as an environmental bath for the single atom. We observe
accelerated decoherence of the single atom when ultracold atoms have
thermalized. This is explained by the emergence of chaotic eigenstates
in the thermalized system. 
\end{abstract}


\maketitle
It is well known that a quantum mechanical state, prepared as a linear
combination of one sort or other, exhibits interference phenomena
according to the rules of quantum mechanics \cite{takagi02}. An orthodox
example is a single particle confined in a double well potential.
The particle wave function can be viewed as a superposition of being
in the left and in the right wells respectively. This fact explains
the coherent oscillations of the particle through the barrier in the
double well. Such quantum mechanical coherence (or interference) has
been shown to be washed out by the influence of the environment and
experiments reveal a spatial localization of the particle in one well. 

The theory of decoherence of open quantum systems is well developed
\cite{schlosshauer}. The mechanism of decoherence is classified into
dissipation and dephasing. To account for dissipation, the environment
is usually represented as an infinite number of harmonic oscillators,
which absorb energy and cause the system to dissipate to the ground
state. The ground state is still coherent, although the particle does
not perform oscillations. Other models \cite{hepp} attempt to explain
dephasing, which turns a pure quantum mechanical state into a mixture.
In the mixed state the particle is found in either well with equal
probability. Decoherence in quantum systems, which are coupled to
external environments, have been observed \cite{exper}.

While dissipation can also be caused by classical means, dephasing
is a purely quantum mechanical effect. A common approach to this problem
is a statistical description based on the density matrix master equation,
which describes an open quantum system in the presence of an environment
\cite{takagi02,schlosshauer}. The rules of quantum mechanics, however,
are applicable to closed quantum systems. Moreover, the unitary nature
of quantum mechanics seems to prevent a system from showing irreversible
loss of coherence. However, the possibility of isolated finite quantum
systems thermalizing to constant values of measurables was theoretically
predicted \cite{srednicli94} and has already been numerically demonstrated
\cite{rigol08,us,Santos}. It has been shown that temperature and
entropy can be assigned to describe such systems \cite{flambaum97}.
Until now, the demonstration that an isolated quantum system could
serve as a finite thermal heat bath for another quantum system was
missing. Here, we achieve this by coupling ultracold atoms confined
to a double well to a single atom. Our analysis shows that this realistic
system exhibits thermalization when one well is initially hotter than
the other \cite{us} and therefore the system can serve as a finite
thermal environment for a single particle in the same double well.
\textcolor{black}{We show that thermalization is a very effective
mechanism in our finite system to cause decoherence of the single
atom. Anything like that is observed for the same numbers of non-thermal
atoms.} We explain this by the emergence of chaotic eigenstates in
the thermalized system.

This also brings the possibility to ascribe a universal wave function
that links environment and objects as parts of a single quantum system.
\textcolor{black}{The need to describe measurements within the quantum
mechanics of closed systems is long-standing and was put forward by
Everett in Ref. \cite{everet}. In this spirit, we can regard tentatively
the thermal environment as an apparatus \cite{zureck91} monitoring
the single atom in this closed quantum system: the single atom becomes
entangled with the apparatus and its state turns to a quantum mixture
of states observers perceive. }

\begin{figure}
\includegraphics[width=8cm]{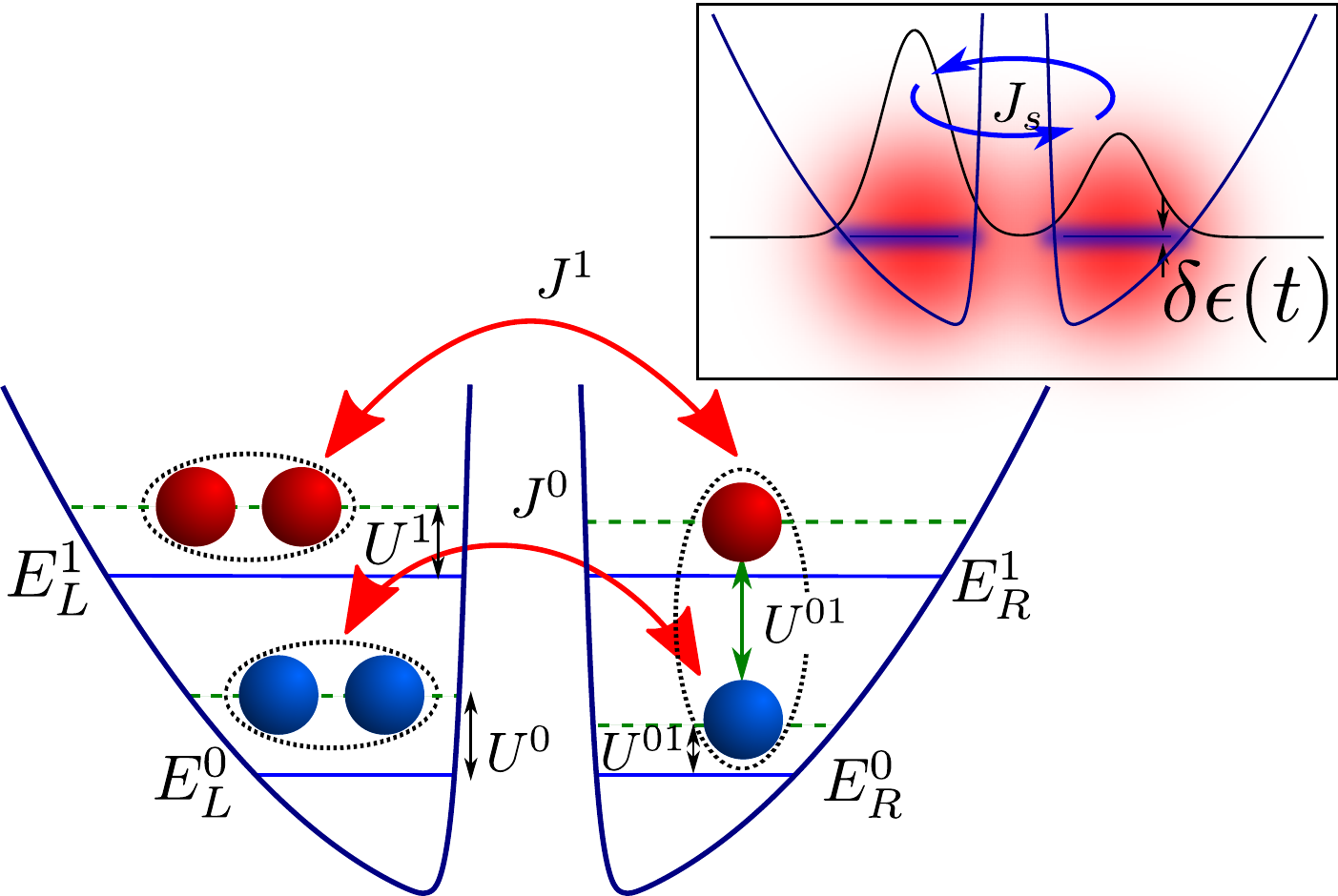}

\caption{Schematic of a double well created by splitting a harmonic potential
with a focused laser. The diagram shows the possible tunneling of
atoms and energy levels change due to interactions. Inset: A single
atom of a different type coupled to them suffers decoherence. The
particle can be found in a well with certain probability.\label{fig:system}}
\end{figure}

We consider $N$ bosons confined to a double well trapping potential,
$V_{dw}(x)$, which are described by the following Hamiltonian \begin{eqnarray} \nonumber \hat{H}_{\rm T}&=& \int dx \left[ \frac{-\hbar^2}{2m} \nabla \hat{\psi}^{\dag}(x) . \nabla \hat{\psi}(x) + V_{dw}(x) \right] \\ &+& g\int dx \hat{\psi}^{\dag}(x)\hat{\psi}^{\dag}(x)\hat{\psi}(x)\hat{\psi}(x). \label{eq:2welsecquant} \end{eqnarray}
At low temperatures only the lowest laying single particle states
are populated. Here we take into account the first two states in each
well as depicted in Fig.~\ref{fig:system}. Therefore the field operators
can be described in terms of the four localized single particle functions,
$\hat{\psi}(x)= \sum_{l=0}^1 \left( \phi_L^{l}(x) \hat{b}_L^{l} + \phi_R^{l}(x) \hat{b}_R^{l} \right)$,
where $\hat{b}_r^l$ are the bosonic annihilation operators of an
atom in well $r$ and energy level $l$ and described by the single
particle functions $\phi_r^l$. This leads to the two-band Hubbard
Hamiltonian \cite{carr} \begin{eqnarray} \nonumber \hat{H}_{\rm T}&=&-\sum_{ r\ne r' ,l} J^l \hat{b}_r^{l\dagger}\hat{b}_{r'}^{l}+\sum_{r,l}U^l\hat{n}_r^l (\hat{n}_r^l-1) +\sum_{r,l}E_r^l \hat{n}_r^l \\ &&+U^{01}\sum_{r,j\ne l'}(2\hat{n}_r^l \hat{n}_r^{l'}+\hat{b}_r^{l\dagger}\hat{b}_{r}^{l\dagger} \hat{b}_r^{l'}\hat{b}_{r}^{l'}), \label{eq:2wel2levelBHM} \end{eqnarray}
where we have ignored interactions between atoms in different wells.
The ground and first excited state energies are $E_r^l = \int dx \phi_r^{l*}(x) \left(- \frac{\hbar^2}{2m}\nabla^2+ V_{dw}(x) \right) \phi_r^{l}(x)$.
The tunnel coupling between the wells are $J^l = \int dx \phi_L^{l*}(x) \left(- \frac{\hbar^2}{2m}\nabla^2+ V_{dw}(x) \right) \phi_R^{l}(x)$.
The interaction between atoms in the same well and on the same energy
level is $U^l = g \int dx |\phi_r^{l}(x)|^4$, and on different energy
levels is $U^{01} = g \int dx |\phi_r^{0}(x)|^2 |\phi_r^{1}(x)|^2$.
This last term also leads to atoms changing energy levels. \textcolor{black}{It
was shown in Ref.\cite{us}}\textcolor{red}{{} }\textcolor{black}{that
this system exhibits thermalization when one well is initially more
energetic than the other. }

We add a single atom of a different atomic species to the system \textcolor{black}{which
is described by} the field operator $\hat{\psi}_{\rm s}$. It interacts
with the thermal atoms via contact interaction $g_{\rm I}\int dx \hat{\psi}^{\dagger}(x)\hat{\psi}^{\dagger}_{\rm s}(x)\hat{\psi}_{\rm s}(x)\hat{\psi}(x)$.
We will assume that it is cold and can be described by the lowest
\textcolor{black}{states of the wells}. Similarly to the previous
case we split the quantum field operator for the single atom as $\hat{\psi}_{\rm s}(x) = \psi_L(x) \hat{a}_L + \psi_R(x) \hat{a}_R$,
where $\hat{a}_r$ are the bosonic annihilation operators of the single
atom in well $r$. This procedure results in the following Hamiltonian
of the combined system \begin{eqnarray}\nonumber \hat{H}&=&\hat{H}_{\rm T}-J_{\rm s} \sum_{r\ne r'}\hat{a}_r^{\dag} \hat{a}_{r'} \\ &+&g_{\rm I}\sum_{l,m=0}^1 \sum_{\alpha,\beta,\gamma,\delta=L}^R C_{\alpha,\beta,\gamma,\delta}^{l,m} \left(\hat{b}_{\alpha}^{l}\right)^{\dag} \hat{b}_{\beta}^{m} \hat{a}_{\gamma}^{\dag} \hat{a}_{\delta}, \label{Eq.Hamilt} \end{eqnarray}
where $J_{\rm s}=\int dx \psi_L^{*}(x) \left(- \frac{\hbar^2}{2m_{\rm s}}\nabla^2+ V^{\rm s}_{dw}(x) \right) \psi_R(x)\sim (E_1-E_0)$
and the interaction coefficients are $C_{\alpha,\beta,\gamma,\delta}^{l,m} = \int dx \phi_{\alpha}^{l\ast} (x)\phi_{\beta}^m (x) \psi^{\ast}_{\gamma} (x) \psi_{\delta} (x)$.
It is interesting to note that the interaction terms induce an effective
tunneling for the single particle ($\gamma\ne\delta$), which is the
result of the shifting of the energy levels due to interaction with
thermal atoms. It also induces on-site interactions ($\gamma=\delta$)
leading to the phase shifts which is assumed by the mechanism of dephasing.
\textcolor{black}{Moreover, each term in the Hamiltonian conserves
the number of atoms in each species.}\textcolor{red}{{} }\textcolor{black}{This
implies that energy dissipation can not occur and that the decoherence
found below is due to dephasing. }

We consider a harmonic potential with oscillator frequency $\omega_0$,
which is split \textcolor{black}{into two halves} by a focused laser
beam and described by a Gaussian potential \textcolor{black}{$V_0\exp[-x^2/2\sigma^2]$.}
The barrier height $V_0 = 10\hbar\omega_0$, with width $\sigma=0.1 l_{ho}$,
where $l_{ho}= \sqrt{\hbar/m\omega_0}$ is the harmonic oscillator
length. For a symmetric well, localized functions representing the
energy levels in the different wells were calculated from the single
particle eigenstates of the system. This gives $J^0/\hbar\omega_0 = 0.153$,
$J^1/\hbar\omega_0 = 0.226$, $E_r^0/\hbar\omega_0 = 1.37$ and $E_r^1/\hbar\omega_ 0= 3.31$.
The interaction terms can be calculated from the integrals above and
the interaction couplings, $g$ and \textcolor{black}{$g_{\rm I}$.}
The interaction couplings can be varied by the Feshbach resonance
\cite{bloch08,feshbach10} and for our purpose we use $U^0/\hbar\omega_0 = 2/N$, $U^1=3U^0/4$ and $U^{01} = U^0/2$.
\textcolor{black}{We assume that the single atom is twice as heavy
as a thermal atom, yielding $J_s/\hbar\omega_0\approx 0.1$, and it
is coupled to thermal atoms with $g_{\rm I}/\hbar\omega_0=2/N$. The
matrix $C_{\alpha,\beta,\gamma,\delta}^{l,m}$ can be calculated in
a similar way. Its most relevant elements will be discussed later.}

The basis are level occupation kets. For thermal atoms it is $|n_L^0,n_L^1,n_R^0,n_R^1\rangle$
with the constraint $n_L^0+n_L^1+n_R^0+n_R^1=N$, \textcolor{black}{where
the total number of thermal atoms $N$ is fixed.} A single particle
can be in the left or in the right well in the second quantized formalism,
therefore the basis for the single particle $|n_L,n_R\rangle$ is
spanned by the two kets, $|1,0\rangle$ and $|0,1\rangle$. The basis
of the full system is then a tensor product of both $|n_L^0,n_L^1,n_R^0,n_R^1\rangle \otimes |n_L,n_R\rangle$.
The initial state can be set as a product state. In the course of
time evolution it is no longer a product state due to interactions
and it becomes an entangled state of the single atom and the thermal
atoms: \begin{eqnarray} |\Psi(t)\rangle &=& \sum_{\{n_r^l\}} A_{\{n_r^l\}}(t)|n_L^0,n_L^1,n_R^0,n_R^1\rangle |1,0\rangle \nonumber \\ &+&\sum_{\{n_r^l\}} B_{\{n_r^l\}}(t)|n_L^0,n_L^1,n_R^0,n_R^1\rangle |0,1\rangle \label{Eq.Psit} \end{eqnarray}
with some time dependent coefficients $A_{\{n_r^l\}}(t)$ and $B_{\{n_r^l\}}(t)$.
The initial state can be chosen such that $A_{\{n_r^l\}}(0)=1$ and
$B_{\{n_r^l\}}(0)=0$ for a particular initial configuration of thermal
atoms $\{n_r^l\}$. 
\begin{figure}
\includegraphics[width=9cm,height=5cm]{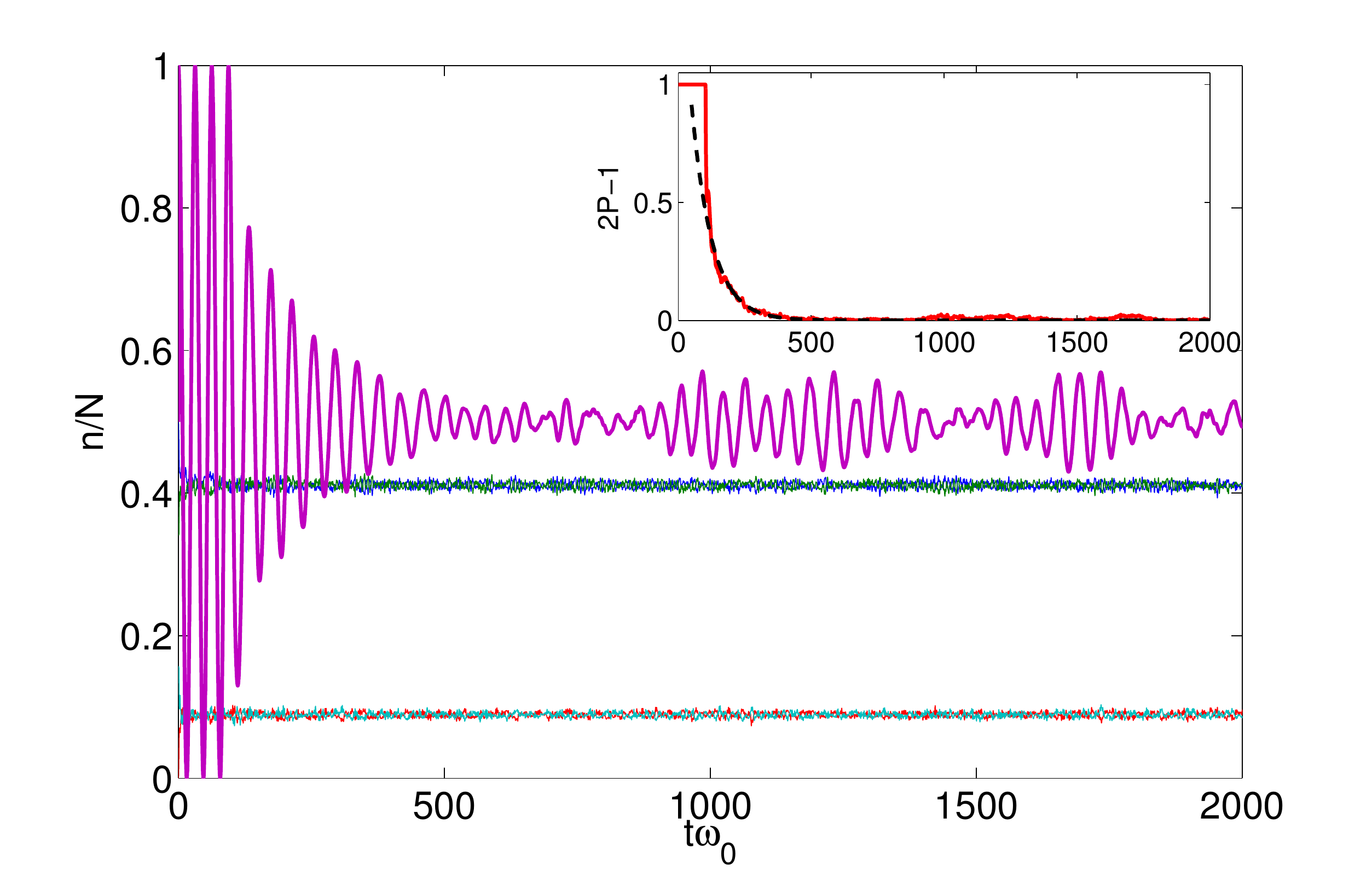}

\caption{Time evolution of the\textcolor{red}{{} }\textcolor{black}{population
expectation values of the energy levels in the system for $N=30$.
The initial state of the thermal atoms is $|16,10,0,4\rangle$ and
the single atom is initially in the left well. Only}\textcolor{red}{{}
}\textcolor{black}{the population of the single atom in the left well
is shown.} Thermal atoms are allowed to thermalize initially. The
coupling between thermal atoms and the single atom, $g_{\rm I}/\hbar\omega_0=2/N$,
is switched on at $t\omega_0=100$. Oscillations of the single atom
amplitude decay with time after that. The corresponding purity of
this state is shown in the inset. The dashed line behaves as \textcolor{black}{$\sim \exp(-0.012t\omega_0)$}.\label{fig:dephasing}}
\end{figure}
 The wave function (\ref{Eq.Psit}) can be viewed as a superposition
of the single atom being in the left and the right well. We can define
a reduced density matrix by tracing the full density matrix over the
thermal atoms, $\hat{\rho}(t)={\rm Tr} \{|\Psi(t)\rangle \langle\Psi(t)|\}$,
which yields \begin{equation} \hat{\rho}(t)=\sum_{\{n_r^l\}}\left(\begin{array}{cc}|A_{\{n_r^l\}}(t)|^2 & A_{\{n_r^l\}}^{\ast}(t)B_{\{n_r^l\}}(t) \\ A_{\{n_r^l\}}(t)B_{\{n_r^l\}}^{\ast}(t) & |B_{\{n_r^l\}}(t)|^2 \end{array}\right) \end{equation}The
purity is defined as $P=\mbox{Tr}\hat{\rho}^{2}$. It is 1 for a pure
state and 0.5 for a decohered state \cite{schlosshauer}.

\begin{figure}
\includegraphics[width=8.5cm,height=5.3cm]{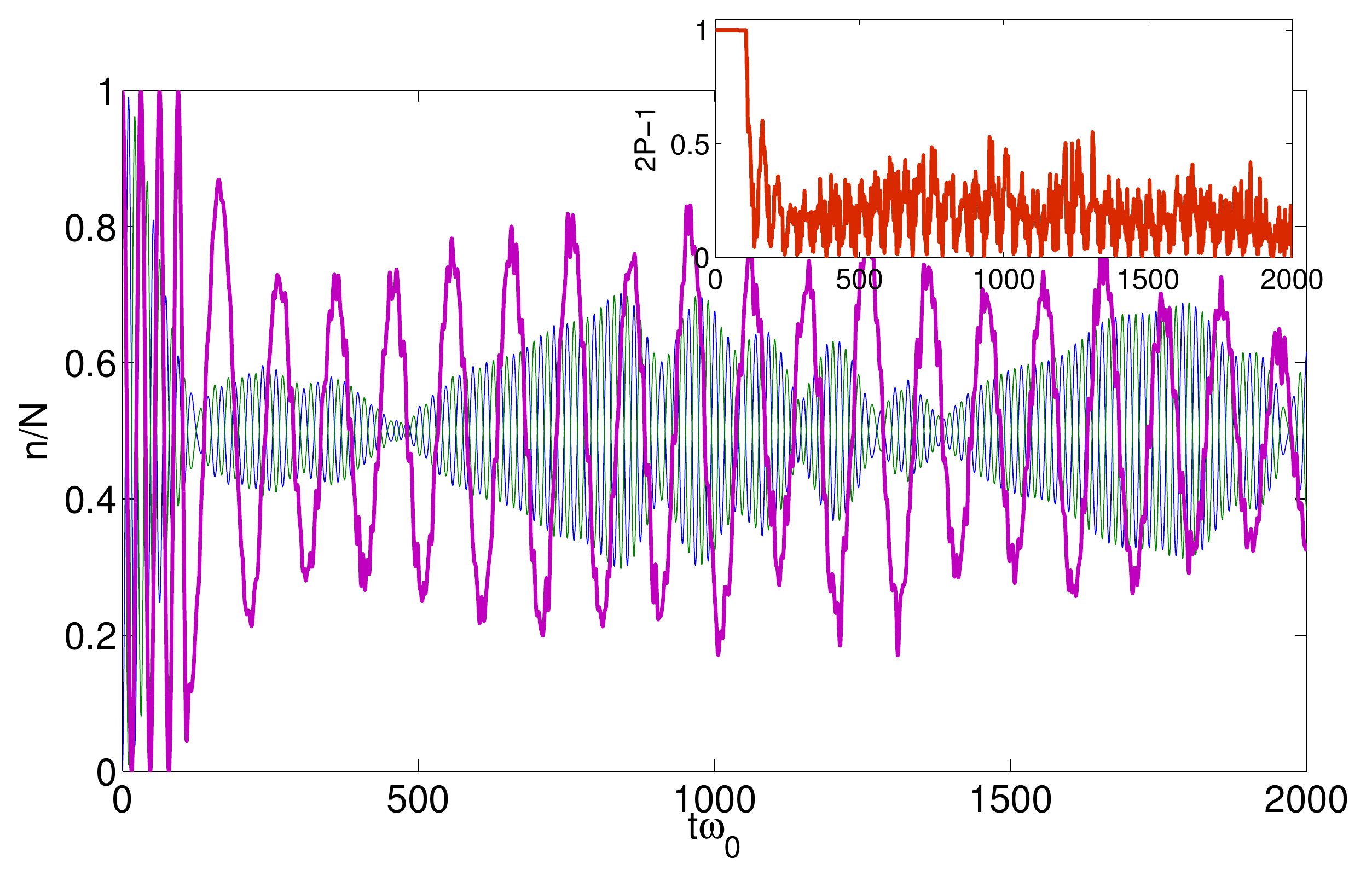}

\caption{\textcolor{black}{Similar to what is shown in Fig. \ref{fig:dephasing},
but now all atoms are allowed to move only in the lowest band. The
initial state of the atoms is $|30,0\rangle$. Contrary to Fig. \ref{fig:dephasing}
the oscillations of the single atom amplitude do not decay and the
corresponding purity exhibits revivals. }\label{fig:distr}}
\end{figure}

The numerical calculations based on the exact diagonalization of the
combined Hamiltonian (\ref{Eq.Hamilt}) reveal dephasing of the coherent
oscillations of the probability of the single atom being found in
one of the wells (see Fig.~\ref{fig:dephasing}). We let the system
thermalize initially. After that we couple the single atom to the
thermalized atoms at $t\omega_0=100$ and observe amplitude decay.
The purity of this state is shown in the inset. The sudden coupling
between the single and the thermal atoms causes the amplitude oscillations
of the single atoms and its purity to drop initially and establish
an exponential decay after that. As anticipated it quickly reaches
the value $0.5$ during the dephasing process, such that $2P-1$ becomes
$0$. \textcolor{black}{We have checked whether the similar behavior
occurs when we switch off the coupling $U^{01}$ in Eq. (\ref{eq:2wel2levelBHM}).
In this case the system is simply reduced to the single-band Hubbard
model which does not show thermalization \cite{walls97}. We perform
numerical simulations for $N=30$ which are allowed to reside only
in the lowest band. To avoid self-trapping regime and allow atoms
to fluctuate we reduce the interaction strength to $U^{0}/\hbar\omega_{0}=0.1/N$
while keeping the coupling to the single atom the same $g_{{\rm {I}}}/\hbar\omega_{0}=2/N$.
As it is seen from Fig. \ref{fig:distr} the single atom oscillations
are not damped away even though it is entangled to a complex environment.
Therefore thermalization is a very effective mechanism in our system
to cause decoherence of the single atom. We do not observe anything
like that for the same numbers of non-thermal atoms. }

In order to understand the observed behavior we first discuss general
arguments leading to decoherence \textcolor{black}{in a thermalized
environment.} Consider a particle confined to a double well potential
and the system is coupled to a thermal environment shown schematically
in Fig. \ref{fig:system}. In the absence of an environment the particle
can perform coherent oscillations from one well to another. If the
two lowest energy states of the system are $|0\rangle$ and $|1\rangle$
respectively, then the linear combinations $|L\rangle = (|0\rangle - |1\rangle )/\sqrt{2}$
and $|R\rangle = (|0\rangle + |1\rangle )/\sqrt{2}$ represent a situation
where the particle is located in the left or the right well respectively.
The state $|L\rangle$ then evolves in time as

\begin{equation}
|\psi(t)\rangle=\frac{1}{\sqrt{2}}[\exp(-iE_{0}t/\hbar)|0\rangle-\exp(-iE_{1}t/\hbar)|1\rangle],\label{eq:1}
\end{equation}
where $E_0$ and $E_1$ are the energies of the two states. The interaction
with an environment\textcolor{red}{{} }\textcolor{black}{leads to energy
level shifts and induces} additional phase factors. As a result, the
phase factors in (\ref{eq:1}) become $E_0't/\hbar+\theta_0(t)$ and
$E_1't/\hbar+\theta_1(t)$. Written in the $|L\rangle$ and $|R\rangle$
basis, the density matrix reads

\begin{equation}
\hat{\rho}(t)=\left(\begin{array}{cc}
\frac{1}{2}[1+\cos{\theta(t)}] & \frac{i}{2}\sin{\theta(t)}\\
-\frac{i}{2}\sin{\theta(t)} & \frac{1}{2}[1-\cos{\theta(t)}]
\end{array}\right),\label{eq:rho}
\end{equation}
where $\theta(t)=(E_1'-E_0')t/\hbar + \theta_1(t)-\theta_0(t)$. \textcolor{black}{If
we consider a thermal heat bath,} the induced phases \textcolor{black}{are
supposed to be} uncorrelated, therefore one may take an appropriate
stochastic average. \textcolor{black}{The process in question is meant
to be also Gaussian implying} $\langle\cos\theta(t)\rangle=\cos(t/T)\exp(-\Theta(t)/2)$
and $\langle\sin\theta(t)\rangle=\sin(t/T)\exp(-\Theta(t)/2)$, where
$T=\hbar/(E_1'-E_0')$ is the period of oscillations and $\Theta(t)=\langle (\theta_1(t)-\theta_0(t))^2\rangle$
describes their damping. \textcolor{black}{In the thermalized environment
we expect the phase to exhibit random walks yielding}\textcolor{red}{{}
}\textcolor{black}{$\Theta(t)\sim t$ }\cite{corrphase}. \textcolor{black}{The}\textcolor{red}{{}
}\textcolor{black}{density matrix (\ref{eq:rho}) thus} quickly becomes
diagonal with time, losing its coherent off-diagonal elements, $\hat{\rho}(\infty)={\rm diag}(1/2,1/2)$.
The purity $P={\rm Tr}\hat{\rho}^2(t)$ changes from 1 to 0.5 during
this transition. Another interesting consequence is that the probability
of finding the particle in one of the wells performs damped oscillations
with period $T$. This analysis assumes that the \textcolor{black}{thermal}
environment is too complex to be considered explicitly and does not
allude to a mechanism of dephasing. Below we will show how random
phases appear in our system and cause decoherence of the single particle.
\begin{figure}
\includegraphics[width=1\columnwidth]{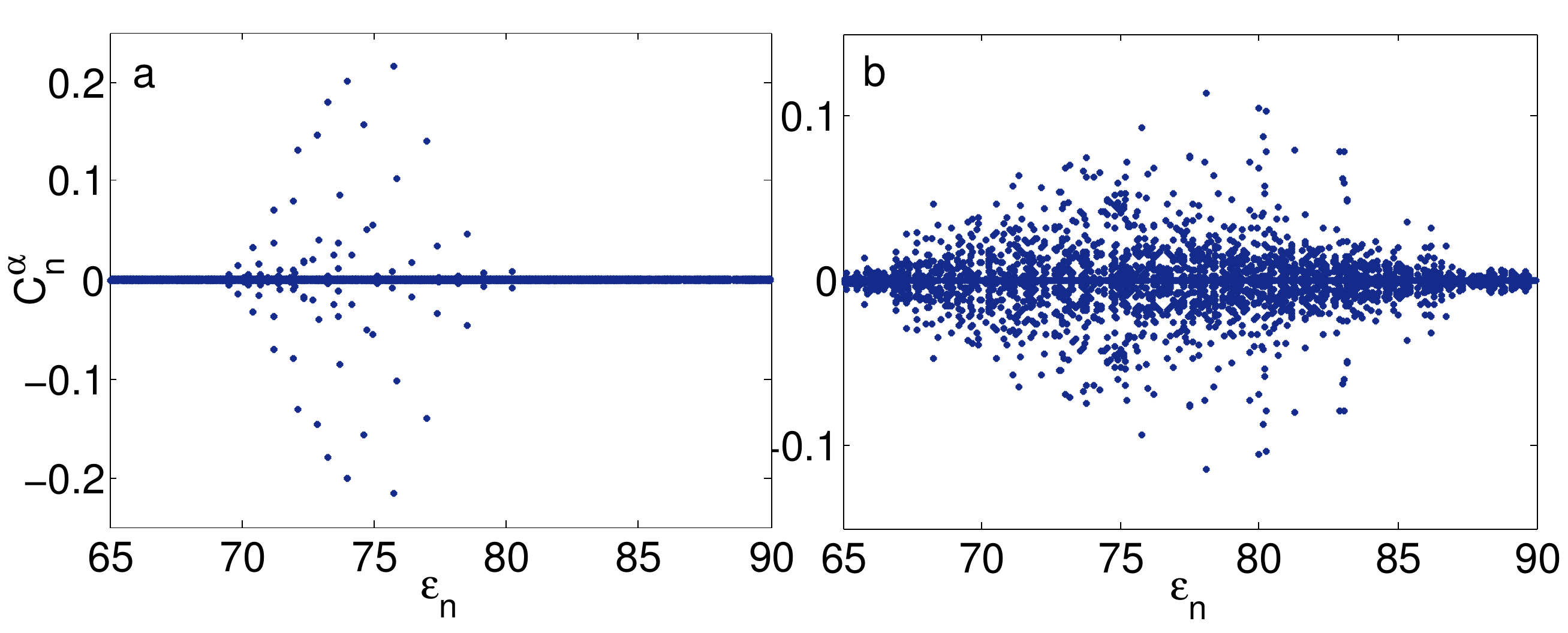}\caption{\label{fig:ChaoticEig} Typical eigenstates of $\hat{H}_{T}$. $\epsilon_{n}=\langle n|\hat{H}_{T}|n\rangle$
are the diagonal elements of $\hat{H}_{T}$ in the fock basis $|n\rangle$.
In panel (a) $U^{01}=0$ and the system does not show thermalization
since it is reduced to the one-band Hubbard model. In panel (b) $U^{01}=\hbar\omega_{0}/N$
is finite and the system thermalizes. In contrast to (a), the eigensates
in (b) are chaotic \cite{Santos}.}
\end{figure}

\begin{figure}
\includegraphics[width=0.95\columnwidth]{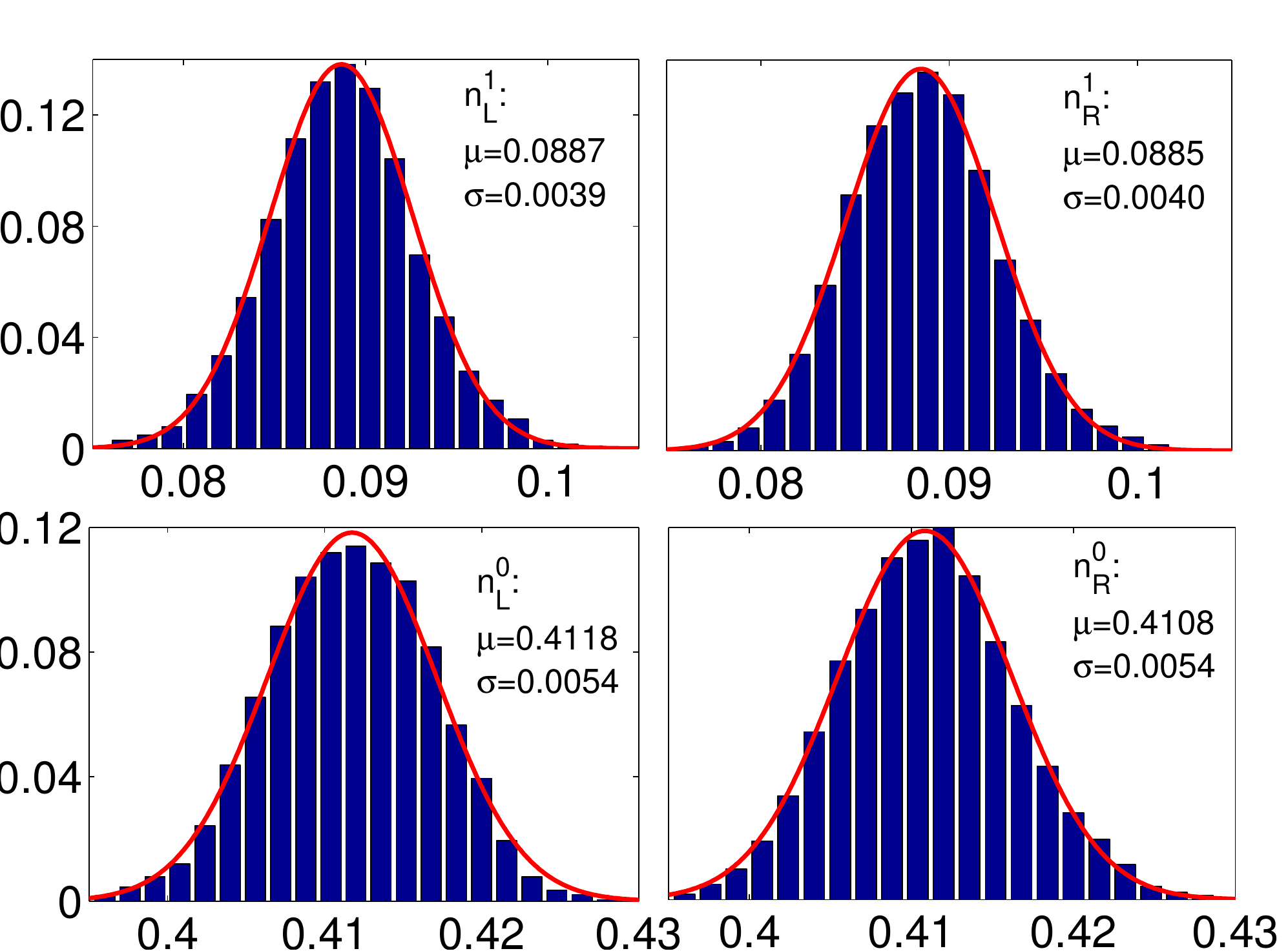}\caption{The probability distributions of the occupations of thermal atoms
for N = 30. They acquire Gaussian shapes for sufficient number of
particles and this serves as evidence of thermalization.\label{fig:The-probability-distributions}}
\end{figure}

Let $|\psi_{T}(t)\rangle$ be the quantum state of the thermalised
system at time $t$ satisfying $|\psi_{T}(0)\rangle=\sum_{\alpha}A_{\alpha}|\alpha\rangle$
and given by the unitary evolution 

\begin{equation}
|\psi_{T}(t)\rangle=e^{-i\hat{H}_{T}t/\hbar}|\psi_{T}(0)\rangle=\sum_{\alpha}A_{\alpha}e^{-iE_{\alpha}t/\hbar}|\alpha\rangle.
\end{equation}
Here $|\alpha\rangle$ is an eigenstate of $\hat{H}_{T}$ given in
Eq.~\ref{eq:2wel2levelBHM} and $E_{\alpha}$ is the corresponding
eigenvalue. The eigenstate can be expanded in terms of the basis kets
$|\alpha\rangle=\sum_{n}C_{n}^{\alpha}|n\rangle$, where we have introduced
a shortened notation $|n\rangle=|n_{L}^{0},n_{L}^{1},n_{R}^{0},n_{R}^{1}\rangle$.
The time-dependent expectation value of the occupation of one of the
levels in the double well then reads

\begin{equation}
\langle\hat{n}_{r}^{l}\rangle=\sum_{\alpha}|A_{\alpha}|^{2}n_{\alpha\alpha}+\sum_{\alpha\ne\beta}A_{\alpha}^{\ast}A_{\beta}e^{i(E_{\alpha}-E_{\beta})t/\hbar}n_{\alpha\beta}.
\end{equation}
The first term on the right-hand side is the long-time average of
$\langle\hat{n}_{r}^{l}\rangle$, the thermalized value to which it
should relax \cite{srednicli94}. The second term represents thermal
fluctuations. We observe that $n_{\alpha\beta}=\sum_{n}nC_{n}^{\alpha\ast}C_{n}^{\beta}$
is a fluctuating quantity with zero mean if $\alpha\ne\beta$$ $.
The reason for that is the emergence of chaotic eigenstates in the
thermalized quantum system \cite{Santos}. It is demonstrated in Fig.\ref{fig:ChaoticEig}.
The width of the fluctuations in Fig. \ref{fig:ChaoticEig},b is $\sim\Delta^{2}/\delta^{2}$,
where $\Delta$ is the mean level spacing of unperturbed system, i.e.
for $U^{01}=0,$ and $\delta^{2}={\cal N}^{-1}\mbox{Tr}V^{2}$ with
$V=U^{01}\sum_{r,j\ne l'}(2\hat{n}_r^l \hat{n}_r^{l'}+\hat{b}_r^{l\dagger}\hat{b}_{r}^{l\dagger} \hat{b}_r^{l'}\hat{b}_{r}^{l'})$
being perturbation and ${\cal N}$ is the size of the Hilbert space.
The width becomes smaller as we increase the number of particles,
since typically $\Delta$ behaves as $\sim e^{-\mbox{const}N}$. As
the result the mean expectation values also fluctuate accordingly.
The probability distribution of the thermal atom density fluctuations
after $t\omega_0=100$\textcolor{black}{{} show clear Gaussian behavior
with the mean values of populations $0.41$ and $ $$0.09$ on the
lower and upper energy levels with the corresponding variances $0.005$
and $0.004$ as shown in Fig.\ref{fig:The-probability-distributions}.}
The width of the Gaussian profile gets narrower as we increase the
total number of particles and behaves as $\sim 1/\sqrt{N}$ in accordance
with the predictions of statistical mechanics \cite{Reif}. \textcolor{black}{Contrary,
the fluctuations become more pronounced as we decrease number of particles.
Thermalization is lost when the number of atoms less than $N\sim10-12$.
The single atom exhibits vivid quantum revivals and decoherence is
no longer observable. }

Now, we can understand the behavior in Fig.~\ref{fig:dephasing} by
analyzing Eq.~(\ref{Eq.Hamilt}). Having discovered the probability
distributions of the thermal atom's density fluctuations, we may substitute
the operators $\hat{b}_r^l$ and $\hat{b}^{l\dagger}_r$ by their
mean-field values and study the remaining quantum problem for the
single atom, which reads \begin{equation} \hat{H}_s\approx -(J_{\rm s}-\Delta J)\sum_{r\ne r'}\hat{a}_r^{\dag} \hat{a}_{r'} + \epsilon_L\hat{a}_L^{\dag} \hat{a}_{L}+ \epsilon_R\hat{a}_R^{\dag} \hat{a}_{R}, \label{Eq.single} \end{equation}where
$\Delta J\approx g_{\rm I}\sum_{l=0}^1 \sum_{\alpha=L}^R C_{\alpha,\alpha,L,R}^{l,l} n_{\alpha}^l$ and $\epsilon_r\approx g_{\rm I}\sum_{l=0}^1 \sum_{\alpha=L}^R C_{\alpha,\alpha,r,r}^{l,l} n_{\alpha}^l$.
Here we have retained the most relevant terms. Notice that the mean-field
values $n_r^l$ are regarded as Gaussian random variables. This also
makes $\Delta J$ and $\epsilon_r$ fluctuating variables with means
$J_0=0.5\times 10^{-2}$, $\epsilon_0=0.12$ and variances $1.5\times 10^{-5}$,
$1.2\times10^{-3}$ respectively. We can safely ignore the fluctuations
of the induced tunneling $\Delta J$ and write $\epsilon_r(t)\approx\epsilon_0+\delta\epsilon(t)$
and $\Delta J(t)\approx J_0$, where $\delta\epsilon(t)$ is a Gaussian
random variable with zero mean. We define a column vector ${\bf A}=(\hat{a}_L, \hat{a}_R)^{\rm T}$.
The Heisenberg equations of motion corresponding to Eq.~(\ref{Eq.single})
are \begin{equation} i\hbar\frac{\partial{\bf A}}{\partial t}=[(\epsilon_0+\delta\epsilon(t))\hat{1}-J'_s\hat{\sigma}_x]{\bf A}, \label{Eq.Heisenberg1} \end{equation}where
$J'_s=J_s-J_0$ is the shifted single particle tunneling rate and
$\hat{\sigma}_i$ are Pauli matrices. The unitary transformation ${\bf A}=\exp[-i(\epsilon_0\hat{1}-J'_s\hat{\sigma}_x)t/\hbar]\tilde{\bf A}$
eliminates the time-independent coefficients in the above equation.
The resulting equation can be solved yielding $\tilde{\bf A}(t)=\exp[-i\int_0^tdt' \delta\epsilon(t')/\hbar]\tilde{\bf A}(0)$.
Averaging over its realizations we get $\tilde{\bf A}(t)\approx\exp[-1/4\Theta(t)]\tilde{\bf A}(0)$,
where \textcolor{black}{the damping} $\Theta(t)\approx 2\hbar^{-2} \int_0^t dt'\int_0^t dt''\overline{\delta \epsilon(t')\delta \epsilon(t'')}.$ Assuming $\overline{\delta \epsilon(t')\delta \epsilon(t'')}=\sigma^2\exp[-2|t'-t''|/\tau_c]$,
where the variance $\sigma=1.2\times 10^{-3}$, we can calculate the
integral $\Theta(t)\approx 2\hbar^{-2}\sigma^2\tau_c t$ which is
valid for $t\gg \tau_c/2$. Here $\tau_c$ is the correlation time
of the Gaussian fluctuations. The time $\tau_c/2$ can thus be regarded
as the time for establishing the exponential decay in Fig.\ref{fig:dephasing}.
Taking for simplicity $\tau_c \sim \hbar/\sigma$ \textcolor{black}{as
the time scale associated with the thermal at}oms fluctuations, we
estimate that $\Theta(t)\approx 2\hbar^{-1}\sigma t\sim 0.01t\omega_0$.
This is in good agreement with the behavior of the dashed line in
Fig.~\ref{fig:dephasing}. We notice that $\Theta(t)\propto g_{\rm I}$.
We checked numerically that this is indeed the case by changing the
interaction strength $g_{\rm I}$.

\textcolor{red}{}

A tight magnetic trap with radial and axial frequencies of $\omega_{\perp}/2\pi=2.1$kHz
and $\omega_{0}/2\pi=11$Hz can be used to confine atoms in a 1D trapping
potential \cite{gring12}. In this case the ratio of the corresponding
oscillator lengths is $l_{ho}/l_{\perp}\approx 14$, such that the
width of the focused laser beam is $\sigma\approx l_{\perp}$. For $^{7}$Li
atoms this gives $\sigma\approx 0.8\mu m$. A narrow laser beam was
recently reported with $\sigma\approx 0.7 \mu m$ and positioning
the beam to a lateral precision of $0.05 \mu m$ \cite{bloch12}.
Taking the 1D interaction strength $g=2\hbar\omega_{\perp} a_s$ yields
the estimate for the two-body scattering length $a_s\sim 200$nm$/N$.
The scattering length for $^7$Li atoms was achieved as small as $\sim 10^{-4}$nm
\cite{hulet09}, therefore the number of particles should not exceed
$N \sim 10^6$. 

In conclusion, we have demonstrated how a quantum particle suffers
decoherence in an isolated thermalized environment of cold bosonic
atoms confined to a double well potential. In particular, we have
shown that the emergence of chaotic eigenstates accelerates the process
of decoherence even if the number of environmental atoms is small.
\textcolor{black}{Hence, our system acts as a finite heat bath for
the particle. The heat bath could potentially be utilized to build
autonomous quantum devices such as refrigerators or engines. }We hope
our work will engage further studies in this direction.

We thank Jacob Dunningham, Ronnie Kosloff and Lukas Gilz for useful
discussions. Special thanks to Dr. David Hallwood for his insights,
comments and help in preparing this manuscript. This work was supported
by the Marsden Fund (contract No. MAU1205), administrated by the Royal
Society of New Zealand.

\end{document}